\def\year{2022}\relax
\documentclass[letterpaper]{article} 
\usepackage{aaai22}  
\usepackage{times}  
\usepackage{helvet}  
\usepackage{courier}  
\usepackage[hyphens]{url}  
\usepackage{graphicx} 
\def\UrlFont{\rm}  
\usepackage{natbib}  
\usepackage{caption} 
\DeclareCaptionStyle{ruled}{labelfont=normalfont,labelsep=colon,strut=off} 
\usepackage{algorithm}
\usepackage{algorithmic}
\usepackage{dsfont}
\usepackage{amsmath}
\usepackage{booktabs}
\usepackage{amssymb}
\usepackage{multirow}
\usepackage[table,xcdraw]{xcolor}
\usepackage{titling}
\usepackage{newfloat}
\usepackage{listings}
\floatstyle{ruled}
\newfloat{listing}{tb}{lst}{}
\floatname{listing}{Listing}

\makeatletter
\newcommand{\smallsym}[2]{#1{\mathpalette\make@small@sym{#2}}}
\newcommand{\make@small@sym}[2]{%
  \vcenter{\hbox{$\m@th\downgrade@style#1#2$}}%
}
\newcommand{\downgrade@style}[1]{%
  \ifx#1\displaystyle\scriptstyle\else
    \ifx#1\textstyle\scriptstyle\else
      \scriptscriptstyle
  \fi\fi
}
\makeatother

\title{High-quality Conversational Systems}
\author{
    Samuel Ackerman,\textsuperscript{\rm 1}
    Ateret Anaby-Tavor, \textsuperscript{\rm 1}
    Eitan Farchi,\textsuperscript{\rm 1}
    Esther Goldbraich,\textsuperscript{\rm 1}
    George Kour,\textsuperscript{\rm 1}
    Ella Rabinovich,\textsuperscript{\rm 1} \\
    Orna Raz, \textsuperscript{\rm 1}  
    Saritha Route,\textsuperscript{\rm 2} 
    Marcel Zalmanovici,\textsuperscript{\rm 1} 
    Naama Zwerdling\textsuperscript{\rm 1}   \\
    }
\date{
    \textsuperscript{\rm 1}IBM Research, Haifa\\
    \textsuperscript{\rm 2}IBM Consulting, India\\
}

\begin{document}
\maketitle

\begin{abstract}
Conversational systems or chatbots are an example of AI-Infused Applications (AIIA). Chatbots are especially important as they are often the first interaction of clients with a business and are the entry point of a business into the AI (Artificial Intelligence) world. The quality of the chatbot is, therefore, key. However, as is the case in general with AIIAs, it is especially challenging to assess and control the quality of chatbot systems. Beyond the inherent statistical nature of these systems, where occasional failure is acceptable, we identify two major challenges. The first is to release an initial system that is of sufficient quality such that humans will interact with it. The second is to maintain the quality, enhance its capabilities, improve it and make necessary adjustments based on changing user requests or drift. These challenges exist because it is impossible to predict the real distribution of user requests and the natural language they will use to express these requests. Moreover, any empirical distribution of requests is likely to change over time. This may be due to periodicity, changing usage, and drift of topics.  

We provide a methodology and set of technologies to address these challenges and to provide automated assistance through a human-in-the-loop approach. We notice that it is crucial to connect between the different phases in the lifecycle of the chatbot development and to make sure it provides its expected business value. For example, that it frees human agents to deal with tasks other than answering human users. Our methodology and technologies apply during chatbot training in the pre-production phase, through to chatbot usage in the field in the post-production phase. They implement the `test first' paradigm by assisting in agile design,  and support continuous integration through actionable insights.

\end{abstract}

\section{Introduction} \label{sec:intro}

Conversational systems or chatbots have become a key channel for customer engagement. Chatbots are automated systems through which users can interact with the business through a natural language interface. For many customers, the chatbot provides their first interaction with the business. Business-grade chatbot systems are tasked to provide the most suitable responses consistently and at the same time provide the best possible customer experience. The chatbot has to be reliable, be consistent, interpret user intents correctly, and respond appropriately. 

\begin{figure}[ht]
\centering
\includegraphics[width=0.45\textwidth]{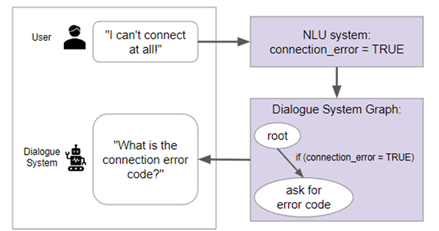}
\caption{Simplified example of a chatbot system. Typically, the system is composed of a learning part tasked with Natural Language Understanding (NLU), usually called intent classifier, and a classical software part, usually a rule-based dialog system that is tasked with handling the conversation flow.}
\label{fig:chatbotHighLevel}
\end{figure}

Figure \ref{fig:chatbotHighLevel} depicts a simplified example of a chatbot system. The system is typically composed of two major parts: 
\begin{enumerate}
    \item a machine learning part (NLU system in Figure \ref{fig:chatbotHighLevel}) that has to correctly classify the user queries called \textit{utterances}, such as the utterance  “I can't connect at all!”,  into their topic classes called \textit{intents}, and 
    \item a dialogue system (Dialogue system graph in Figure \ref{fig:chatbotHighLevel}) usually implemented as a rule-based graph, potentially with some learning elements, that has to navigate the entire conversation to its  successful completion.
\end{enumerate}
 A successful completion is one in which the user either gets the information or actions that they requested, or is promptly directed to a human agent. In many customer care scenarios if the chatbot system is able to identify that it is having challenges in providing a good response to the user, the  conversation will be handed off to a human agent to resolve the case or a ticket will be raised to have a human agent look into the problem and resolve it. An unsuccessful completion or an unhandled conversation is one in which the user abandons the conversation or is otherwise unhappy with the response they get. 
The automation via the chatbot system needs to create a positive and rich customer experience, and enable serving of relevant content and returning customers, be able to ‘understand’ the customer need as a human would, and respond accordingly. 

The quality of the chatbot needs to be sufficient even when first released. However, often at this stage, representative data may not be available, as it is almost impossible to manually predict the way people will approach the chatbot.
Moreover, even if the first chatbot release is of satisfactory quality, the chatbot implementation will need to be updated to handle new situations. These may be due to changes or drift in the distribution of user utterances, including new topics, periodicity in topics and expiration of topics. The chatbot needs to maintain and improve its quality, a requirement that is especially challenging in light of the changing interactions.
Unfortunately, chatbots may fail to provide a high-quality customer experience because they do not understand the customer’s intent, are not designed to cover enough situations, or even fail to respond appropriately to the user request and loop back to repetitive interactions.
A high quality chatbot should handle drift gracefully. For example, drifted calls should be routed quickly to a human agent and their data collected systematically from the agent to help define new topics, update the training data, update the test data, or otherwise refine the design of the chatbot system.

We identify three major challenges to chatbot quality:
\begin{enumerate}
    \item A first release of a chatbot system that is of sufficient quality that the users are willing to interact with. Section \ref{subsec:preprod} provides details.
    \item Growing the chatbot system based on human-to-bot logs, while maintaining and improving the quality of the system. Sections \ref{subsec:postprod},\ref{subsec:grow} provide details.
    \item Identifying and handling drift. Section \ref{subsec:grow} provides details.
\end{enumerate}

We introduce a methodology that is supported by a set of technologies for handling these challenges. 
Our techniques apply throughout the life-cycle of the chatbot system, from pre-production, where overcoming the first challenge is key, to the field or post-production and over time, where overcoming the second challenge is key and the third challenge of drift creeps in. The methodology is agile, advocating fast and continuous feedback between  the pre-production and post-production stages, through actionable insights. This enables continuous integration of the updates to the chatbot systems, through continuous testing and improvement. Our technologies follow the `experiment first' approach \cite{bookExperimentFirst} -- any recommended changes to the chatbot system are a result of running experiments and assessing their results. 
Our methodology requires and assists in defining Key Performance Indicators (KPIs) to ensure that the chatbot system delivers on the business goals.

\section{Methodology} 
\label{sec:methodology}
Testing a chatbot involves interacting with the chatbot and validating its responses to a series of utterances and continued dialogues to determine if the chatbot provides the right responses. This has to be done repeatedly and across all the expected topics or intents to establish consistency of response and coverage of the business intents or functions.  Doing so  requires large volumes of data (or utterances) of a variety of types to statistically test dialogue paths once the intent is identified and also requires data that will uncover irregularities and inconsistencies in the chatbot response.  

Testing should cover all possible dialogue flows, cover ambiguity in response or in intent identification, and cover sensitivities of the machine learning model.  Testing has to be done at various stages of the chatbot development and also periodically when the system is in production. Periodic testing is needed because the chatbot system must `grow', where growth means the process of updating the chatbot system's set of intents. One of the actions required to handle drift is to gradually and continuously add new intents and potentially remove or archive existing intents. Testing assists in ensuring that such changes improve the system or at least do not harm the chatbot system's ability to correctly handle intents it has already been trained to handle and tested on.

Even if we test the chatbot system regularly, it is difficult to establish consistency in the tests, generate large volumes of data, decipher the results of the tests,  and quantify the business value of the chatbot system in its entirety. 

We propose a business metrics-driven approach to test and validate the chatbot system.  Our methodology takes a top-down approach to align chatbot testing with business outcomes. 

Our methodology and techniques:
\begin{itemize}
    \item Define business and technical KPIs, and the means to obtain them. Following are the standard KPIs that we recommend. Of course, the user may define customized KPIs or prioritize among the ones given below.
    \begin{itemize}
        \item Time that the chatbot system saved across human agents in the call center.
        \item Number and percentage of successfully completed, i.e., \textit{contained} calls. 
        \item Number and percentage of calls that the chatbot handles when no human is available  (outside of business hours). 
        \item These become contributing factors to the key KPI of \textit{NPS (Net Promoter Score)} or the user satisfaction from the chatbot. 
        \item \textit{Chatbot coverage} or the number and percentage of intents that the chatbot possess out of all the seen topics. 
        \item User experience  as defined by the number and percentage of effective flows of the conversation such that the user either gets the correct response or is referred to a human agent in a timely manner with a limited number of chatbot interactions.
    \end{itemize}
    \item Apply statistical control on top of the KPIs to decide when the system is no longer operating at the desired level. The methodology achieves this through drift detection and mitigation. 
    \item Provide guidance for identifying weaknesses and sensitivities that are often due to design issues such as poor separation between intents, particularly when these intents are manually labeled.  
    \item Provide a means for communication between the AI Tester who tests the chatbot and the subject matter expert (SME) who trains the chatbot.
    \item Provide  insights that can be acted upon by the tester or the conversational specialist, the data scientist, and SME to improve the system.
    \item Provide significant automation to achieve all the above. For example, provide technologies for generating data, for defining the space to cover in testing, for identifying weaknesses and sensitivities, and for suggesting  potential design issues that might be the cause these.
 \end{itemize}

\begin{figure}[ht]
\centering
\includegraphics[width=0.45\textwidth]{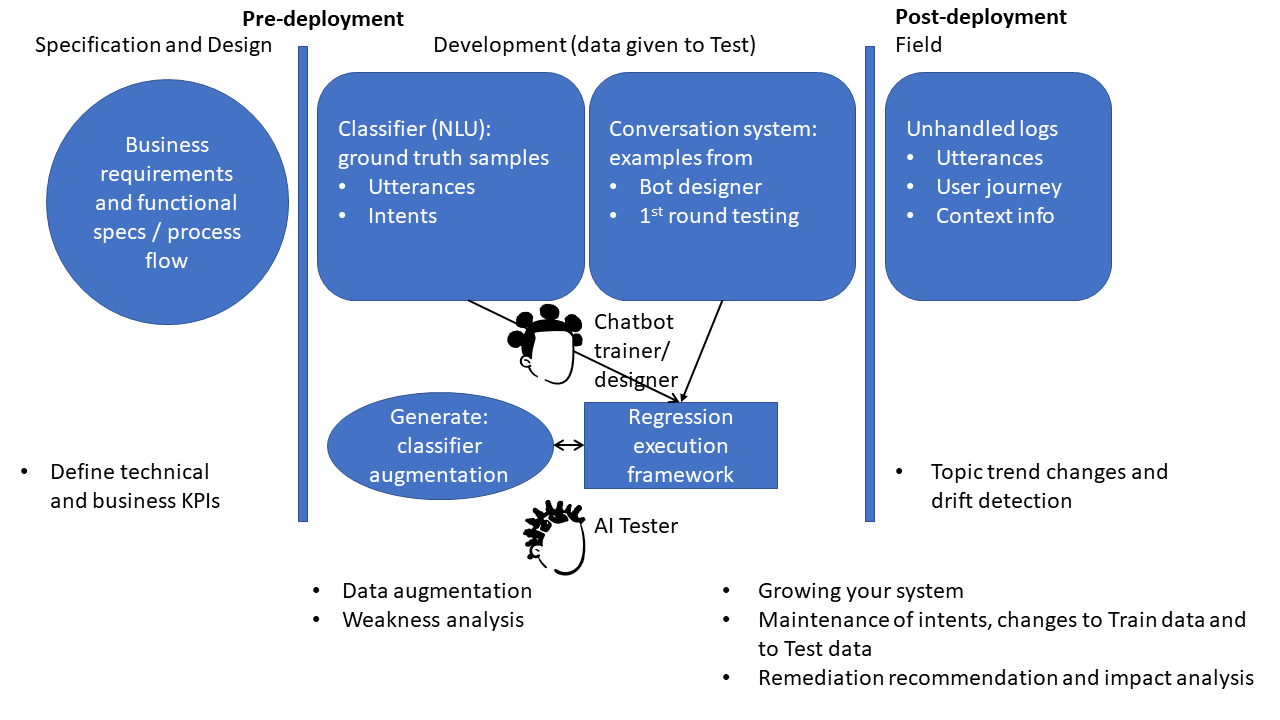}
\caption{Methodology at a glance. Pre-deployment and post-deployment end-to-end chatbot system lifecycle. Top level shapes indicate data created or available at that stage. The chatbot trainer or designer and the AI tester provide inputs and feedback to one another in all phases.}
\label{fig:methodologySummary}
\end{figure}

Figure \ref{fig:methodologySummary} illustrates at a glance the main responsibilities of the chatbot trainer or designer and the AI tester and ther resulting artifacts and data. These persona collaborate closely to test and improve the system continuously. 

\subsection{In pre-production} \label{subsec:preprod}
In pre-production, the methodology defines template data coverage models. These coverage models are designed to test the system's response to varied data. The coverage models reuse the well established technique of Combinatorial Testing (CT).  These input coverage models assist in verifying that the chatbot system will be able to reasonably handle inputs that it was not trained on but are in-domain, and inputs that are especially challenging to classify correctly due to different factors, such as language variations, ambiguous utterances and more. Handling out-of-domain inputs is addressed mainly in the post-production phase via \textit{trend or drift detection and analysis}. Trend changes or drift refer to changes in the topics raised by the humans interacting with the chatbot system. These changes include new topics, change in the way humans articulate existing topics, change in the frequencies of the topics raised, and periodicity. We use the terms trend and drift interchangeably in this paper. 

\subsubsection{CT (Combinatorial Testing)}
 CT is a well-known testing technique to achieve a small and effective test plan. CT is based on the empirical observation that in most cases, the occurrence of a bug depends on the interaction between a small number of features of
the Software Under Test (SUT)~\cite{Bell:2006:OEE,Bell:2005,Kuhn02aninvestigation,Kuhn:2004,Wallace01failuremodes,Zhang:2011}. For example, according to~\cite{Kuhn:2004}, all possible pairs of parameter
values can typically detect 47\% to 97\% of the bugs in an SUT. CT translates this observation to a coverage criterion as follows. It requires a manual definition of the test space in the form of a {\it combinatorial model}, consisting of a
set of parameters, their respective values, and constraints on the value combinations. A valid test in the test space is defined to be an assignment of one value to each parameter that satisfies the constraints. A CT algorithm automatically constructs a subset of the set of valid tests, termed a {\it test plan}, which covers all valid value combinations of every $t$
parameters, where $t$ is usually a user input. Such a test plan is said to achieve 100\% {\it t-way interaction coverage}. 

The following input coverage models automate the task of CT model definition for chatbot testing. Our technologies provide automation support for  test cases creation, as tests translate to utterances and their expected label that comply with  natural language variations or ML model boundaries that the template CT models define.

\subsubsection{Input coverage model 1} Natural language variations. This coverage model defines variations that exist in natural language and may affect any NLP (Natural Language Processing) classifier. These include simple natural language constructs such as contractions and spelling mistakes.

\subsubsection{Input coverage model 2} Natural language classifier decision boundaries.
The coverage model defines characteristics that affect the ability of the classifier to separate among different intents.
One such characteristic is captured by our novel \textit{complexity measure} \cite{complexity}. This measure is independent of the classifier implementation and can be used on the representation resulting from sentence embedding. The complexity measure provides an indication of how difficult it is for any classifier implementation  to classify a given utterance into its correct intent when provided with its ground truth intent and with other utterances and their ground truth intents. Difficulties often stem from design issues, such as defining overlapping intents, ignoring hierarchies that exist within the intents, and intents which are too “wide” or too “narrow” in scope.
Our technology identifies utterances that lie on the classifier decision boundaries or are ambiguous. 
 
 \subsubsection{Data generation}
Data that is representative of the typical way humans will address the  chatbot is key to statistical learning, and the chatbot system intent classifier is often implemented with statistical learning models. 

Testing of 
a chatbot system, which has statistical learning models, means that the AI tester assesses the ability of the system to generalize or to perform well, by outputting mostly correct classification, on inputs that were not seen during training. To do that, there must be sufficient data in testing for statistically-sound analysis. 
However, often there is insufficient data, especially for testing. The LAMBADA technology \cite{lambada} can generate valid and representative textual data per label, given a seed of labeled sentences. The generated data may be used either for training or for testing.

When generating test data the AI Tester will have to assess the validity of the data and its labels. Our complexity measure can be utilized to alleviate some of the human effort needed for validating the generated data. We suggest data buckets, potentially the complexity-based slices that FreaAI finds, of complexity measure levels. Low-complexity utterances are expected to have correct labels with high probability and, therefore, can be assessed with light random sampling for human inspection. High-complexity utterances might be ambiguous and their labels might be incorrect. Therefore, that data would need human inspection.

\subsubsection{Analysis of weaknesses}
The CT input coverage models define what the minimal coverage requirements are for any classifier that works with natural language data. FreaAI weakness analysis \cite{FreaAI1,FreaAI2}  complements these coverage requirements and works on the same input space of meta features (e.g., natural language variations, complexity) but does the analysis directly on the actual values, whether continuous or discrete, without requiring discretization. FreaAI automatically finds statistically significant error concentrations and highlights the meta features ranges and values that are associated with these weaknesses. Therefore, the input CT models and the FreaAI weakness analysis provide different benefits.

The above mentioned input coverage models may fail to find weaknesses that result in error concentrations because these models naively discretize the test or train data. Our data slicing technology FreaAI can automatically find those feature ranges and/or values which result in weak model performance and unexpectedly high error concentration. The template input coverage model may be adjusted using discretization suggested by FreaAI. However, even in that case not all the inputs  will have problematic ranges or statistically significant ones. FreaAI only reports on those ranges and values that are problematic and are highly unlikely to be so by chance. 

\subsubsection{Automation support}
 The definition of the input coverage models is automated and is based on analyzing the given data (usually the test data).
As data is often scarce, particularly in the early or pre- production stages, we provide technology (LAMBADA) that can generate realistic data given a seed of samples per intent.
The analysis of weaknesses (FreaAI) is automated. Weaknesses are analyzed with respect to the factors or meta features that are used in the CT input coverage models.
The analysis of sensitivities is automated and is done per natural language variations.
The analysis of the given data properties is automated.
 
\subsubsection{Reporting and potential remediation} Our technology provides insights about the data and the ML model. These insights often expose design issues and do so while suggesting actions to improve the system. The reporting targets either the AI Tester or the chatbot trainer to facilitate communication of issues and to ease the process of debugging and improving the system. 
 
 We note that there may be different remediation suggestions. These include suggesting enhanced design, updating the training data, and updating the test data.
 
 An example of enhancing design is to add stages to the data processing pipeline prior to the ML model classification. There are natural language constructs that may best be addressed prior to any classification. These include, for example, contractions, case, and spelling.  Contractions may be turned into their underlying full words. Case may be made all lower unless there is a meaning to case variations. Any input could benefit from being first  spell-checked while allowing the user to make changes until correct.

\subsection{In post-production} \label{subsec:postprod}
In post-production the methodology addresses deviations of topics and of language in human-to-bot interactions as compared to those used when training or when testing the chatbot system. When the chatbot system is used in the field there exist additional data in the form of logs recording the interaction of human-to-bot.
Our techniques use these data to continuously improve the chatbot system and to continuously update the chatbot regression tests and training data.
The analysis of logs includes techniques for:
\begin{enumerate}
    \item \textbf{Trend analysis of new topics and detection of drift in existing intents’ data distribution}. Section \ref{subsec:grow} that follows provides more details. The drift detection is statistically sound and provides statistical guarantees on the error types of the alerts. 
    In order to analyze drift and trend changes appropriately there is a need to first establish a baseline. Significant deviations from the baseline will then result in drift alerts. The baseline can be created either in the pre-production phase or in the post-production phase using verified logs.
    Drift detection invokes any of the following suggested actions. These actions include changes to the chatbot system design by changing its Train data, and updates to the Test data to reflect these design changes as well as to relfect usage changes. 
    In addition, we apply drift detection to measurements of the relevant business and technical KPIs that Section \ref{sec:methodology} describes.
    \item\textbf{Changes to the Train and Test data}.
    \begin{enumerate}
        \item  \label{item:rec1} Recommendation of what new utterances to label for training and for dynamic test sets.
        \item \label{item:rec2}  Recommendation for updating the training set schema such as: merge intents, split intents, new intents, archive intents, remove intents. and for updating the test set to reflect these updates. 
    \end{enumerate}
    \item \textbf{Impact analysis.} Analysis of the expected effect or impact of the changes previously described in items \ref{item:rec1} and \ref{item:rec2} on the chatbot classifier performance. 
    \item \textbf{KPI divergence alerts}. Statistically-sound indications when key KPIs such as percentage of contained calls are out of control. This can also be used to modify the business process, e.g., change the number of human agents that are answering uncontained calls.
 \end{enumerate}

 \subsection{Growing your system: Intent maintenance and drift detection} \label{subsec:grow}
 We identify the need for end-to-end, both pre-production and post-production, technology support for adjusting to changes in the topics of user queries, in the way users express their queries, and in the distribution of topics. This is relevant to both first release (pre-production) and on-going releases (post-production). Changes may be due to drift in topics, for example, in the CQA COVID-19 chatbot data, the topics change as the pandemic progresses. At the beginning of the pandemic there may be a lot of questions about quarantine and restrictions, which become less frequent as vaccines become available. New topics appear around the first shot of the vaccines, boosters, vaccines for the elderly, for grownups, and for children. These may appear periodically, or may be replaced by new topics.
 
 In general, after deployment of the initial chatbot, one should expect two aspects of topical trend changes or drift, and be able to address them through our analysis and recommendations.
\begin{enumerate}
     \item New topics that the chatbot designer has not planned for the chatbot to deal with, but still emerge, and stay consistent over time. These may result in adding new intents and actions to handle them.
\item New ways people express their needs and requests within an existing intent. These may result in failure to detect the right intent, and therefore a need for the chatbot designer to update the Intent's examples with a richer set of sentences that better match the actual users utterances. 
 \end{enumerate}

 Dynamically adjusting the conversation system confidence and dialogue flow is a common approach for adjusting to changes. However, that helps in alleviating the consequences of these changes, but does  not address the root cause.
 Our technologies provide a framework where the chatbot trainer and designer can automatically get indications of changes, suggestions for remediation, and what-if impact analysis. 
 
 The actions that the chatbot designer may explore or take, potentially as a result of topic and drift detection include: 
 \begin{itemize}
     \item Relabel existing classifier samples.
     \item Add new samples to an existing intent.
     \item Remove samples from an existing intent.
     \item Add a new intent.
     \item Archive/Remove an existing intent.
     \item Split intents.
     \item Merge intents.
 \end{itemize}
 
 The above actions may be suggested by our techniques following the user-in-the-loop confirmation of the following automatically identified issues:
 \begin{itemize}
     \item Wrong ground truth label or intent.
     \item Hierarchy of labels seems to exist but is implicit. An example of a hierarchy of labels is having `Vaccine' be a top node in the hierarchy, and `first\_shot', `second\_shot', `booster' its children. This 'hidden' hierarchy should be made explicit, even if outside of the classifier. An explicit hierarchy could assist in the labeling effort as well as potentially in designing the conversation flow.
     \item Confused intents -- may be due to labeling that is too fine-grained (then action should be merge), too general (then action should be split), or cannot be separated because the examples are insufficient (add to Train data).  
 \end{itemize}
 Section \ref{subsec:results} that follows provides examples. 
 We have a large body of prior work on drift detection \cite{FreaAIDrift1,FreaAIDrift2,FreaAIDrift3,FreaAIDrift4,FreaAIDriftDensity} and topic trend changes \cite{Ellaswork}. In future work we will put these together to improve the automated detection to topic trend changes and drift, and to create a framework for allowing the user to maintain the data and explore suggested remediation actions.

\section{Evaluation} \label{sec:running_example}
We demonstrate our methodology and technologies on two  publicly available datasets: \textit{CQA},  a COVID-19 Questions and Answers chatbot data \cite{tepper-etal-2020-balancing} and \textit{banking77} a banking related queries chatbot data \cite{Casanueva2020}.

We concentrate on the pre-production phase that Section \ref{subsec:preprod} describes, and demonstrate the effectiveness of our input coverage models, data generation, and weakness analysis. In our analysis we show results of the two input coverage models combined -- including both natural language variations and our complexity measure. 
Future work will concentrate on the post-production and drift detection and mitigation phases that Sections \ref{subsec:postprod} and \ref{subsec:grow} describe. 

The results indicate how our methodology and technologies can assist the AI tester and the Chatbot trainer or conversational specialist to identify design issues. Moreover, our technology can assist the Chatbot trainer to overcome these issues.

\subsection{Data} \label{subsec:example_data}
We provide some characteristic of the datasets that we used in our experiments. These are real datasets and are challenging. For example, as is usually the case with chatbot data, there are dozens of labels for each dataset, the data is imbalanced and has varying numbers of examples per the different classes, and intent names are sometimes confusing.

\begin{figure}[ht]
\centering
\includegraphics[width=0.45\textwidth]{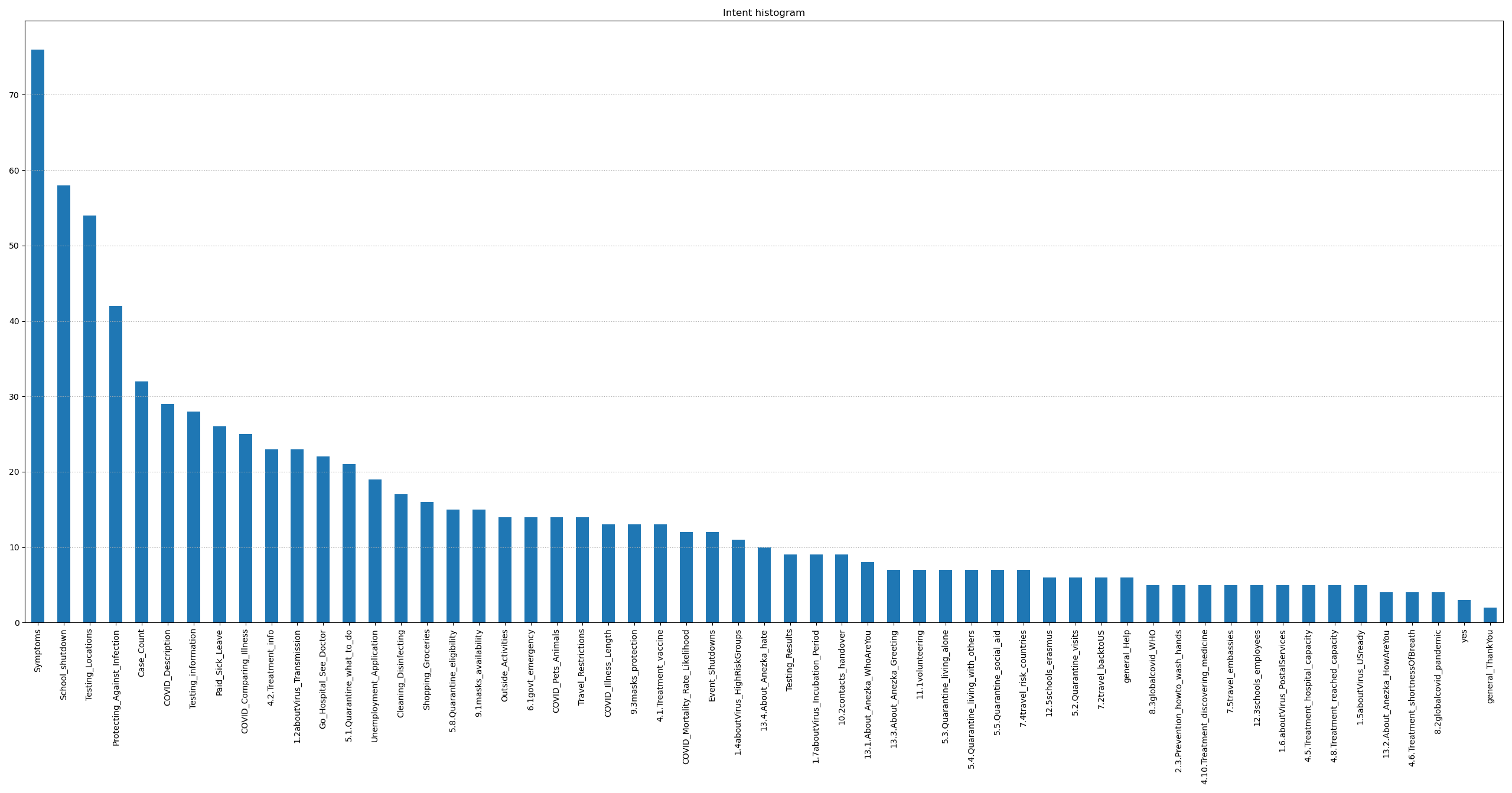}
\caption{Distribution of data records to classes for the CQA dataset via a histogram of examples (utterances) per class (intent).}
\label{fig:CQA}
\end{figure}

\begin{figure}[ht]
\centering
\includegraphics[width=0.45\textwidth]{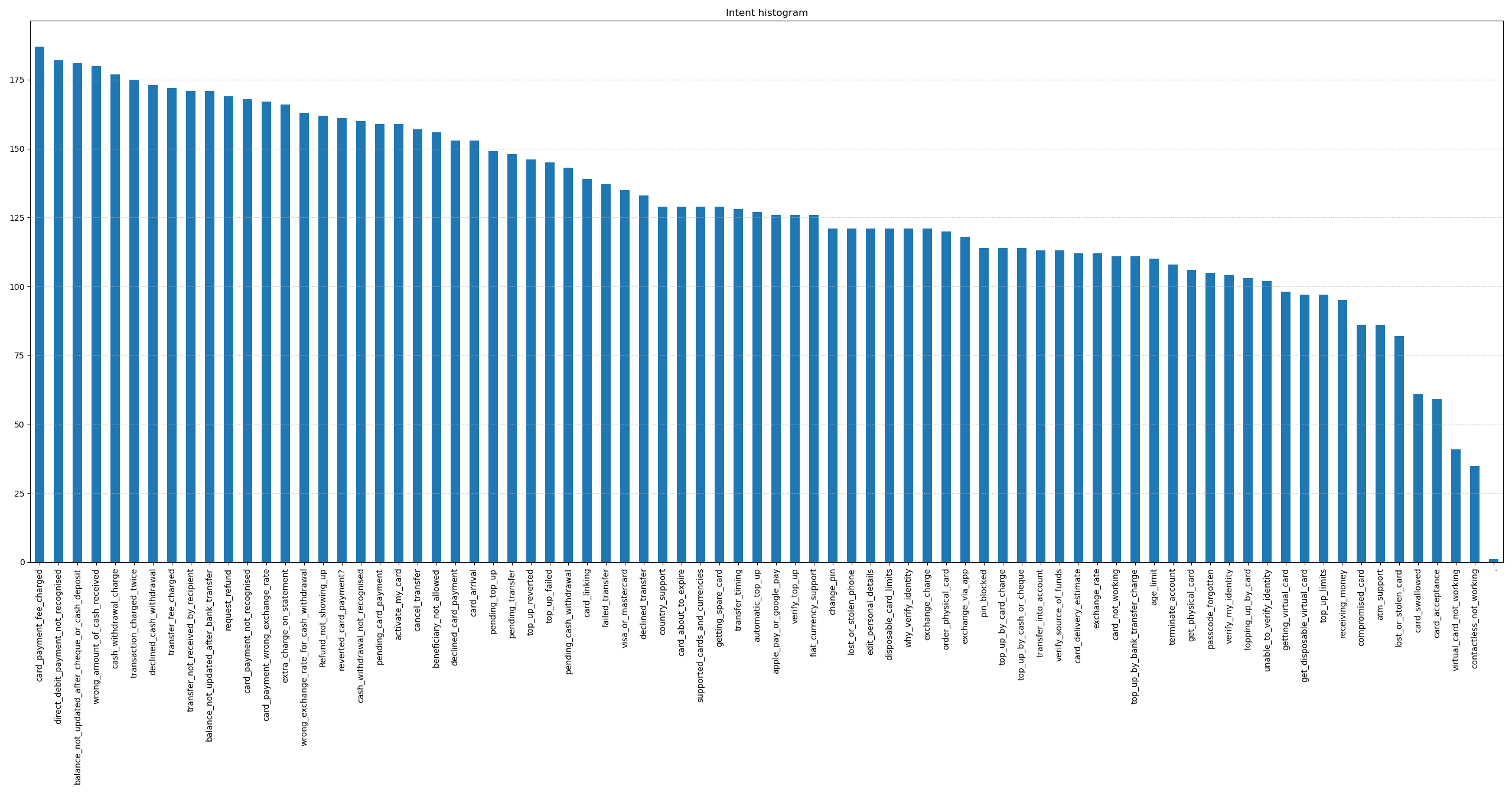}
\caption{Distribution of data records to classes for the banking77 dataset via a histogram of examples (utterances) per each of its 77 classes (intents).}
\label{fig:banking77}
\end{figure}

Figures \ref{fig:CQA} and \ref{fig:banking77} depict for each class or intent in the CQA and banking77 datasets, respectively, the number of examples per that intent in the Train dataset. We clearly see that the datasets have a skewed distribution of labels, i.e., there are labels with few examples making learning more difficult for such intents.

For our experiments we trained an industrial standard chatbot over the Train set and used it as the chatbot under test for each of CQA and Banking77. We created the Test and Train datasets as follows: we ignored classes with less than 10 examples. For all other classes we divided the data into 80\% Train and 20\% Test datasets. This still resulted in training a chatbot implementation for a decent number of classes (50 classes for CQA, 76 for Banking77), as is usually the case for operational chatbot implementations. 

In our experiments we analyzed both the Test data that contained examples from the original open dataset that we created as described above, as well as on data that was generated using LAMBADA, as data augmentation is part of our methodology. We provided seed data for LAMBADA from the original data, with 5 randomly selected examples per intent. We ran LAMBADA to generate 30 new examples per seed intent.

\subsection{Results highlights} \label{subsec:results}
Our methodology starts with template coverage models for natural language variations and for ML model decision boundaries as Section \ref{sec:methodology} describes. As part of our methodology, in parallel to the pre-defined models variations and their feature boundaries, our technology FreaAI automatically finds weaknesses and the feature boundaries that characterize these weaknesses.

Our methodology also augments  Test data utilizing data generation technology (LAMBADA). For validating this part of our methodology in our experiments, we also created large Test data that was not used for training from the original data, as Section \ref{subsec:example_data} above describes. Indeed, this empirically validated that for testing, the generated data achieves similar results as non-generated data.
The results over the original Test dataset and over the generated Test dataset are similar, providing anecdotal evidence of the ability to utilize generated realistic data for testing. Data augmentation is especially important when test data is scarce. This is often the case in reality. 

Our results indicate that the input coverage models with their pre-defined feature bins or boundaries might miss weaknesses or highlight multiple features as potentially responsible for weaknesses. FreaAI alleviates this issue as its analyses find the interesting feature boundaries while filtering out any values that result in weaknesses that are statistically insignificant, as defined by a hypergeometric distribution. Our complexity measure is consistently associated with statistically significant weaknesses. This result validates the complexity measure, as this measure should capture the inherent difficulty in correctly classifying a sentence into its correct label, given the other sentences in the data set and their labels. This measure is also independent of the actual classifier implementation.
We show the results on data augmented with LAMBADA. The Test dataset that was created from the original open data often had very few instances per variation, such as only 4 instances with all upper case for the banking77 Test data and only 11 instances with a single spelling mistake in the CQA data. This stresses the need to augment the data, even when Test data is generously available as was the case in our experiments, as the likelihood of having all the natural language variations and complexity values naturally occurring is small.

Table \ref{tab:bakingGenCTD} shows the results of computing accuracy and balanced-accuracy for the banking77 generated data. \textit{Balanced-accuracy} is defined as the average of recall (intuitively, the ability of the classifier to find all the positive samples) obtained on each class and is often used when classes are imbalanced (not of equal quantity per label). 
The average accuracy over the entire generated test data is 0.85. Though there are some potential weakness (highlighted in red in the table), these are not severe and it is unclear what to look at first. Table \ref{tab:bakingGenFrea} shows the results of running the FreaAI technology over the actual meta features space -- without pre-binning, as is done for the coverage models defined in Table \ref{tab:bakingGenCTD}. The classifier confidence was added as another meta feature to the FreaAI input in order to ensure FreaAI finds the obvious expected weakness of low accuracy when the classifier confidence is low. FreaAI finds weaknesses that are all associated with complexity values that are relatively high. By the nature of the FreaAI analysis, it only outputs weaknesses if these are statistically significant. This gives a good starting point for investigating the weaknesses.

\begin{table}[] 
\begin{tabular}{llll}
\multicolumn{1}{c}{\textbf{MetaFeature}} & \multicolumn{1}{c}{\textbf{SIZE}} & \multicolumn{1}{c}{\textbf{ACC}}                    & \multicolumn{1}{c}{\textbf{BACC}}                   \\
SPELLING=MULTIPLE\_ERRORS                  & 99                                & \cellcolor[HTML]{FFC7CE}{\color[HTML]{9C0006} 0.78} & 0.82                                                \\
LENGTH=HIGH\_75\_100                       & 558                               & \cellcolor[HTML]{FFC7CE}{\color[HTML]{9C0006} 0.79} & \cellcolor[HTML]{FFC7CE}{\color[HTML]{9C0006} 0.80} \\
CONTRACTIONS=True                          & 332                               & \cellcolor[HTML]{FFC7CE}{\color[HTML]{9C0006} 0.80} & 0.81                                                \\
COMPLEXITY=HIGH\_75\_100                   & 558                               & 0.83                                                & 0.85                                                \\
SPELLING=SINGLE\_ERROR                     & 541                               & 0.83                                                & 0.81                                                \\
COMPLEXITY=MED\_25\_75                     & 1115                              & 0.83                                                & 0.83                                                \\
CASE=ALL\_LOWER                            & 2231                              & 0.85                                                & 0.85                                                \\
CONTRACTIONS=False                         & 1899                              & 0.86                                                & 0.86                                                \\
SPELLING=CORRECT                           & 1591                              & 0.87                                                & 0.86                                                \\
LENGTH=MED\_25\_75                         & 1115                              & 0.87                                                & 0.86                                                \\
LENGTH=LOW\_0\_25                          & 558                               & 0.89                                                & 0.89                                                \\
COMPLEXITY=LOW\_0\_25                      & 558                               & 0.92                                                & 0.88                                               
\end{tabular}
\caption{Input coverage models results of computing accuracy (ACC) and balanced accuracy (BACC) over instances generated for the baking77 data. Each meta feature variation (cell in the first column) had SIZE instances. The numbers in the names of the meta features (first column) indicate the data percentiles that were taken as High (75--100), Medium (25--75), and Low (0--25). \\
The average accuracy over this entire generated Test dataset as well as the average balanced accuracy was 0.85. Cells with a red background indicate a drop in accuracy that may be of interest.}
\label{tab:bakingGenCTD}
\end{table}

\begin{table}[]
\begin{tabular}{llll}
\multicolumn{1}{c}{\textbf{MetaFeature}} & \multicolumn{1}{c}{\textbf{VALUE}} & \multicolumn{1}{c}{\textbf{ACC}} & \multicolumn{1}{c}{\textbf{SIZE}} \\
CONF                                     & 0.227-0.684                        & \cellcolor[HTML]{E6B8B7}0.63     & 700                                  \\
COMPLEXITY                               & 1.487-1.586                        & \cellcolor[HTML]{E6B8B7}0.69     & 55                                   \\
COMPLEXITY                               & 4.012-4.749                        & \cellcolor[HTML]{E6B8B7}0.69     & 35                                   \\
COMPLEXITY                               & 2.446-2.491                        & \cellcolor[HTML]{E6B8B7}0.63     & 30                                  
\end{tabular}
\caption{FreaAI results over instances generated for the baking77 data. The results highlight weaknesses in meta feature actual ranges (VALUE). The weaknesses are all statistically significant, resulting in a drop in accuracy (ACC) over SIZE records as compared to the average accuracy (0.85 in this case) that is highly unlikely to occur by chance. CONF is the classifier confidence. As expected, low confidence often results in an incorrect classification. COMPLEXITY ranges are high and show a correlation with incorrect predictions. This may happen even when the classifier's confidence is high, though that is uncommon. }
\label{tab:bakingGenFrea}
\end{table}

Table \ref{tab:CQAGenCTD} shows the results of computing accuracy and balanced-accuracy for the CQA generated data. The average accuracy over the entire generated test data is 0.78. Though there are some potential weakness (highlighted in red in the table), these are not severe and it is unclear what to look at first. Table \ref{tab:CQAGenFrea} shows the results of running the FreaAI technology over the actual meta features space -- without pre-binning, as is done for the input coverage model defined in Table \ref{tab:CQAGenCTD}. The classifier confidence was added as another meta feature to ensure FreaAI finds the obvious expected weakness of low accuracy when the classifier confidence is low. As was the case for the banking77 dataset, FreaAI finds weaknesses that are all associated with complexity values that are relatively high. This gives a good starting point for investigating the weaknesses.
Moreover, the weaknesses that FreaAI finds have higher error concentration than those of the pre-defined bins. Therefore, they may provide more hints as to the underlying issues. In any case, one needs to further investigate the weaknesses. Our technologies include actionable visualizations to assist a human, usually the chatbot trainer, in this investigation.

\begin{table}[]
\begin{tabular}{llll}
\multicolumn{1}{c}{\textbf{MetaFeature}} & \multicolumn{1}{c}{\textbf{SIZE}} & \multicolumn{1}{c}{\textbf{ACC}}                    & \multicolumn{1}{c}{\textbf{BACC}}                   \\
SPELLING=MULTIPLE\_ERRORS                & 13                                & \cellcolor[HTML]{FFC7CE}{\color[HTML]{9C0006} 0.62} & \cellcolor[HTML]{FFC7CE}{\color[HTML]{9C0006} 0.67} \\
COMPLEXITY=HIGH\_75\_100                 & 160                               & \cellcolor[HTML]{FFC7CE}{\color[HTML]{9C0006} 0.63} & \cellcolor[HTML]{FFC7CE}{\color[HTML]{9C0006} 0.69} \\
LENGTH=HIGH\_75\_100                     & 160                               & 0.76                                                & 0.80                                                \\
SPELLING=SINGLE\_ERROR                   & 129                               & 0.77                                                & 0.72                                                \\
CONTRACTIONS=False                       & 608                               & 0.78                                                & 0.75                                                \\
LENGTH=MED\_25\_75                       & 320                               & 0.78                                                & 0.75                                                \\
CASE=ALL\_LOWER                          & 641                               & 0.78                                                & 0.75                                                \\
SPELLING=CORRECT                         & 499                               & 0.79                                                & 0.74                                                \\
CONTRACTIONS=True                        & 33                                & 0.79                                                & 0.74                                                \\
COMPLEXITY=MED\_25\_75                   & 320                               & 0.79                                                & 0.73                                                \\
LENGTH=LOW\_0\_25                        & 161                               & 0.80                                                & 0.76                                                \\
COMPLEXITY=LOW\_0\_25                    & 161                               & 0.91                                                & 0.83                                               
\end{tabular}
\caption{Input coverage models results of computing accuracy (ACC) and balanced accuracy (BACC) over instances generated for the CQA data. Each meta feature variation (cell in the first column) had SIZE instances. The numbers in the names of the meta features (first column) indicate the data percentiles that were taken as High (75--100), Medium (25--75), and Low (0--25). \\
The average accuracy over this entire generated Test dataset was 0.78. The average balanced accuracy was 0.75. Cells with a red background indicate a drop in accuracy that may be of interest.}
\label{tab:CQAGenCTD}
\end{table}

\begin{table}[]
\begin{tabular}{llll}
\multicolumn{1}{c}{\textbf{MetaFeature}} & \multicolumn{1}{c}{\textbf{VALUE}} & \multicolumn{1}{c}{\textbf{ACC}} & \multicolumn{1}{c}{\textbf{SIZE}} \\
CONF                                     & 0.265-0.556                        & \cellcolor[HTML]{E6B8B7}0.40     & 168                               \\
COMPLEXITY                               & 2.573-2.930                        & \cellcolor[HTML]{E6B8B7}0.50     & 40                                \\
COMPLEXITY                               & 5.009-12.114                       & \cellcolor[HTML]{E6B8B7}0.49     & 39                                \\
COMPLEXITY                               & 1.189-1.311                        & \cellcolor[HTML]{E6B8B7}0.43     & 14                               
\end{tabular}
\caption{FreaAI results over instances generated for the CQA data. The results highlight weaknesses in meta feature actual ranges (VALUE). The weaknesses are all statistically significant, resulting in a drop in accuracy (ACC) over SIZE records as compared to the average accuracy (0.78 in this case) that is highly unlikely to occur by chance. CONF is the classifier confidence. As expected, low confidence often results in an incorrect classification. COMPLEXITY ranges are high and show a correlation with incorrect predictions. This may happen even when the classifier's confidence is high, though that is uncommon. }
\label{tab:CQAGenFrea}
\end{table}

Our analyses can provide similar information for interactions of two or more meta features and their values. However, here we already see problems within a single meta feature. Therefore, we continue with the single meta feature analysis and do not further investigate two and potentially more meta features interacting. 

Zooming in on complexity, we get insights through visualizations of the complexity of intents, in a view of two-dimensional clusters and in a view of a heatmap of selected intents. See examples in Tables \ref{tab:banking77_sentence_examples} and \ref{tab:citizen_assist_sentence_examples} below.

\begin{figure*}[ht]
\centering
\includegraphics[width=0.9\textwidth]{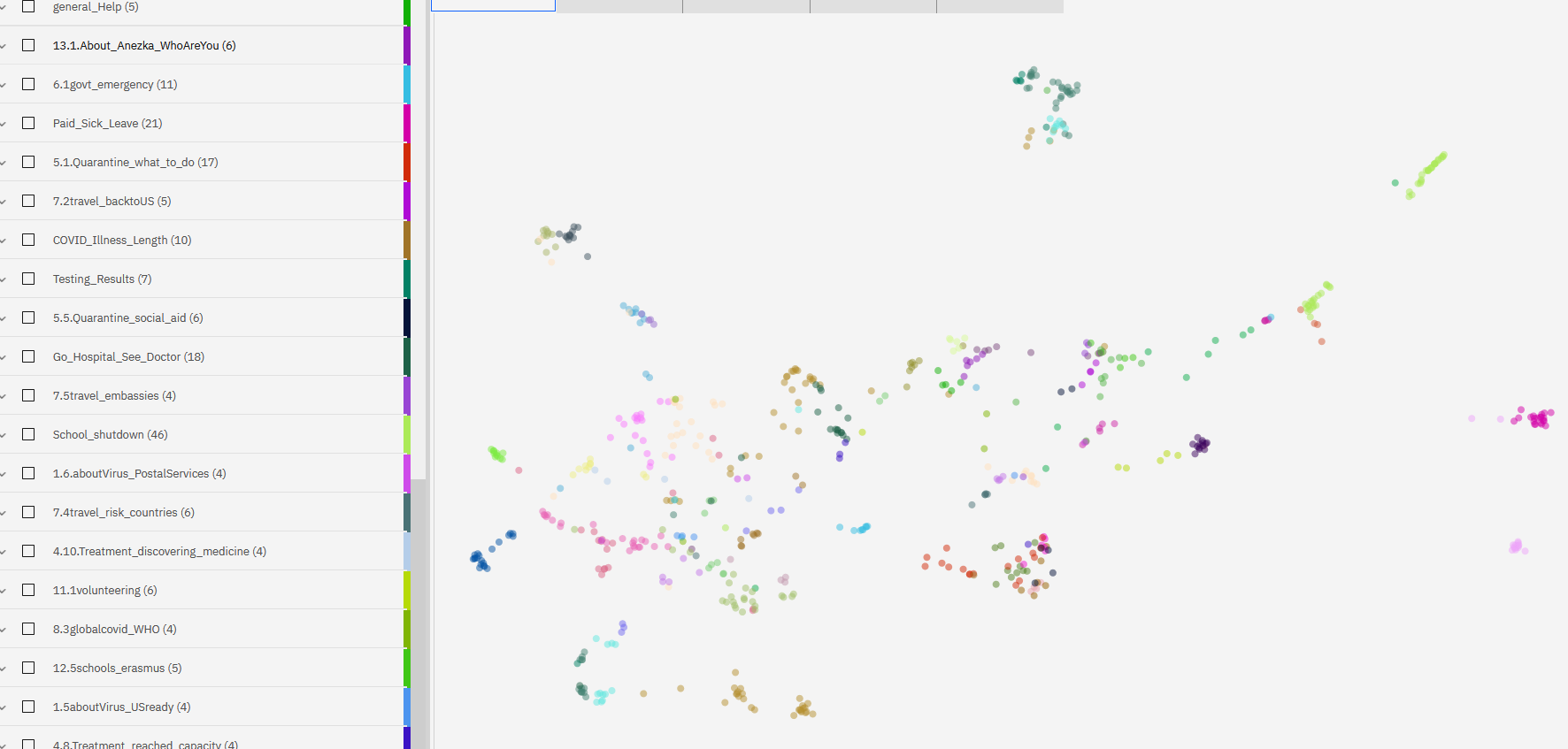}
\caption{CQA data depicted in a 2-D semantic space. Utterances of each intent get the same color}
\label{fig:CQAScatter}
\end{figure*}

Figure \ref{fig:banking77Scatter} shows a visualization of banking77 intents and where in the semantic space their utterances fall. Each utterance is depicted as a point in a 2-D reduction of the original high-dimensionality semantic space. Notice that the complexity computation is done on the original high-dimensional space of the sentence embedding. Similarly, Figure \ref{fig:CQAScatter} shows the semantic space view for the CQA data.

\begin{figure}[ht]
\centering
\includegraphics[width=0.45\textwidth]{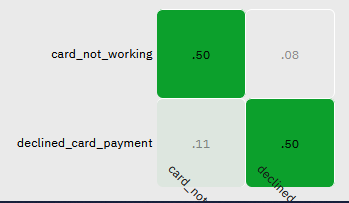}
\caption{Zooming into the complexity between two intents in the banking77 dataset: `card\_not\_working' and `declined\_card\_payment' we see the complexity is non zero. This means that it may be difficult to correctly classify instances due to confusion between these classes. }
\label{fig:banking77heatmap}
\end{figure}

\begin{figure}[ht]
\centering
\includegraphics[width=0.45\textwidth]{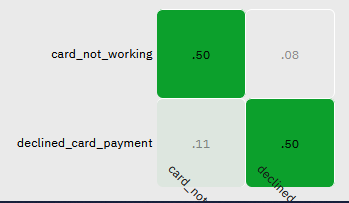}
\caption{Zooming into the two banking77 intents identified in Figure\ref{fig:banking77heatmap}: `card\_not\_working' (pink) and `declined\_card\_payment' (black). Each point represents one sentence that is part of the training data. The same issue that our techniques detected on the generated Test data, as exemplified in the first row of Table \ref{tab:banking77_sentence_examples} repeats in the training data. The highlighted example seems to be incorrectly labeled. It represents the sentence `My card was declined at a restaurant today and I am not sure why' that has the ground truth label `card\_not\_working'. We see that this pink instance is in the midst of the black instances, so it should have probably been labelled black (`declined\_card\_payment'). Moreover, we see a lot of overlap between the pink and the black instances, confirming the class confusion.}
\label{fig:banking77scatter}
\end{figure}

\begin{figure*}[ht]
\centering
\includegraphics[width=0.9\textwidth]{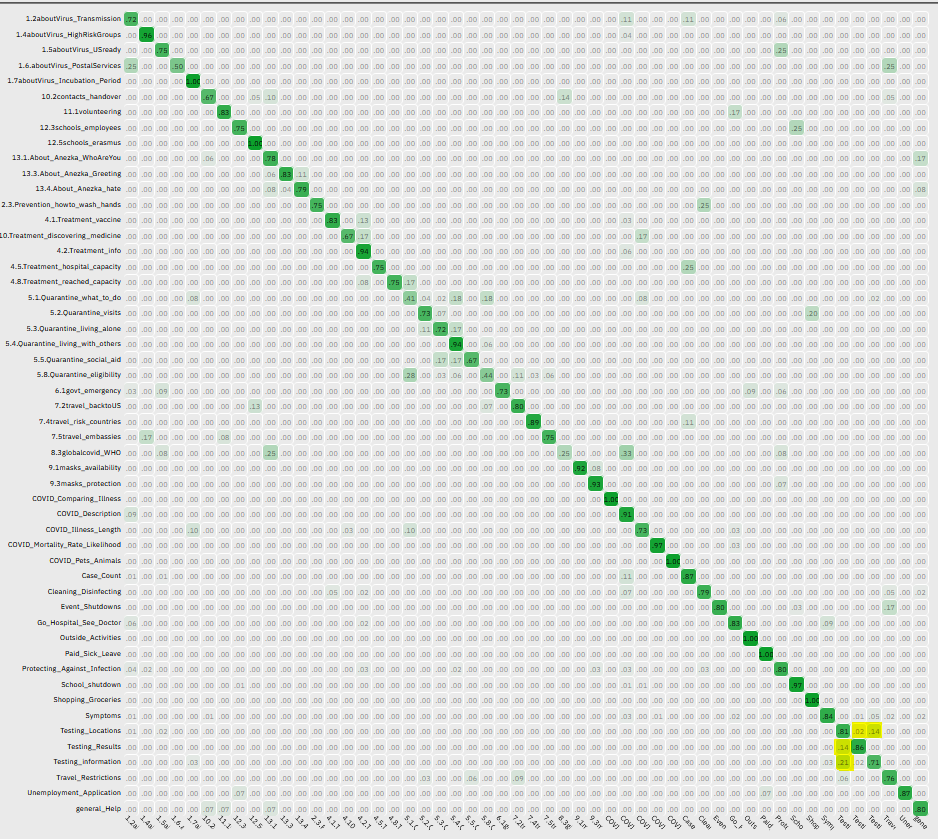}
\caption{CQA heatmap of intent confusion as highlighted by the complexity measure. Cells that are non-zero are confused. Naturally, an intent is most difficult to separate from itself. The yellow highlighted cells are `Testing\_Results', `Testing\_Locations', and `Testing\_Information'. These are difficult to separate. `Symptoms' is another intent that is difficult to separate from these highlighted intents.}
\label{fig:CQAConfusion}
\end{figure*}

Figure \ref{fig:CQAConfusion} shows a visualization of the CQA intents and their complexity in a heatmap view.
Figure \ref{fig:banking77heatmap} shows an example of concentrating the analysis on two of the banking77 intents, based on their complexity measure. These intents are card\_not\_working and declined\_card\_payment. These intents might get confused as indicated by a complexity score that is nonzero. To further investigate this, we look at the scatter plot for these two intents as Figure \ref{fig:banking77scatter} shows. Indeed, there are multiple issues in the training data related to these two intents. The yellow highlighted example is `My card was declined at a restaurant
today and I am not sure why’. It has the ground truth label `card\_not\_working' colored in pink. However, we see that this pink instance is in the midst of the black `card\_payment' instances. The black label may be more appropriate and that instance should probably have
been labelled black. In addition, there is overlap between the pink and the black instances. This may require actions such as providing additional training examples to enable better separation, or merging these two intents, and/or potentially handling them better in the conversation flow. 

\begin{figure*}[ht]
\centering
\includegraphics[width=0.9\textwidth]{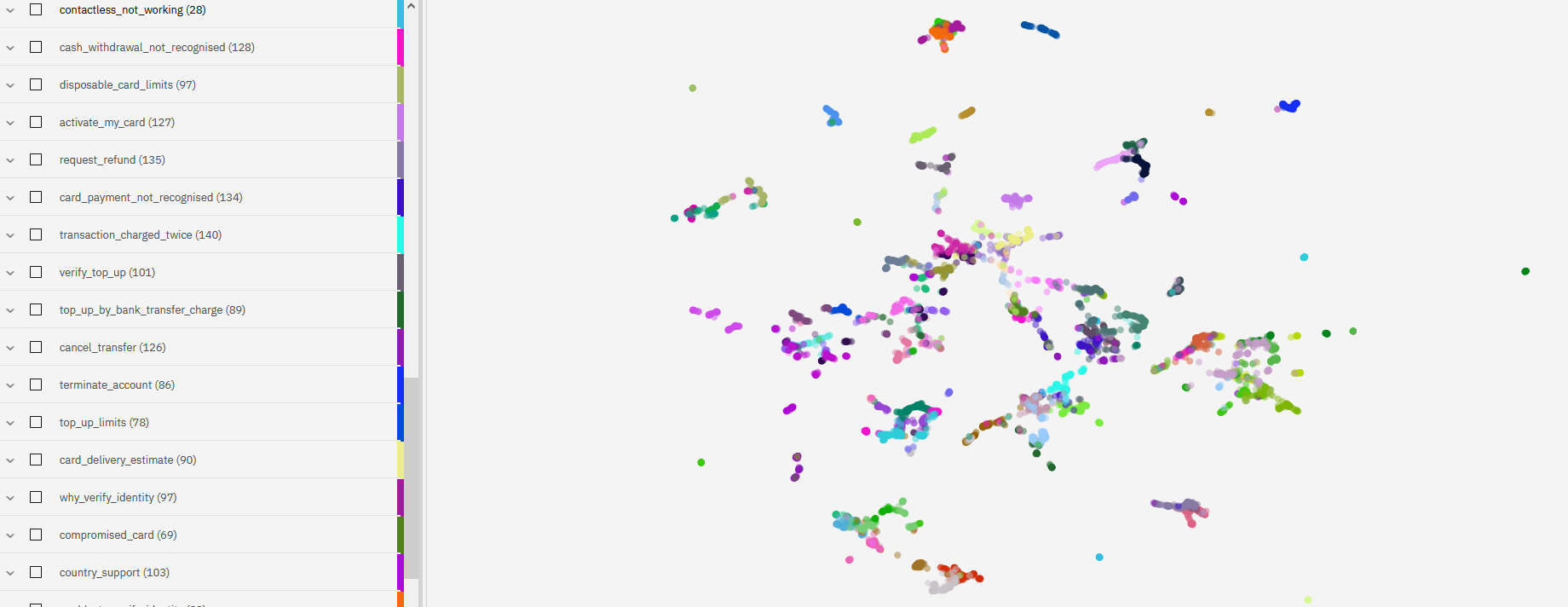}
\caption{Banking77 data depicted in a 2-D semantic space. Utterances of each intent get the same color}
\label{fig:banking77Scatter}
\end{figure*}

As is often the case, we identify label or intent names as an issue. The intent name does not describe well enough the content of utterances with that label. This is a design issue that causes many problems. First, a human will have difficulties assigning the correct intent if the intent name does not capture well its expected topic. Second, many algorithms and heuristics rely on the intent name. For example, a common chatbot flow design pattern is to ask a clarification question based on the intents when a given user utterance is ambiguous. A related pitfall is to print out the intent name as is. This might cause multiple issues. The name may only be clear to an expert but not to the end user, for example if it contains domain-specific synonyms. The name may contain non-words and be inappropriate to expose. The name may not reflect the intent topic and therefore be confusing also to the end user if exposed in that way. 

Our actionable insights can suggest a better fitting name through novel clustering and topic identification technology \cite{Ellaswork}. Moreover, the identified topics may suggest further design issues. These include overlapping intents, intents that are spread over too many topics, and more.

Of course, these intent design issues may exist independent of any naming issues. Additional design issues that may exist, and can be handled with our actionable insights, include having hierarchies of intents but not handling those properly in the Chatbot system implementation or at the very least in the label-assignment process.

\begin{figure}[ht]
\centering
\includegraphics[width=0.45\textwidth]{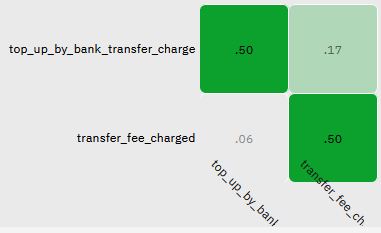}
\caption{Zooming into the complexity between two intents in the banking77 dataset:topup\_transfer\_charge and transfer\_fee\_charged we see the complexity is non zero. This means that it may be difficult to correctly classify instances due to confusion between these classes. }
\label{fig:banking77with2intentscomplexityheatmap}
\end{figure}

\begin{figure}[ht]
\centering
\includegraphics[width=0.45\textwidth]{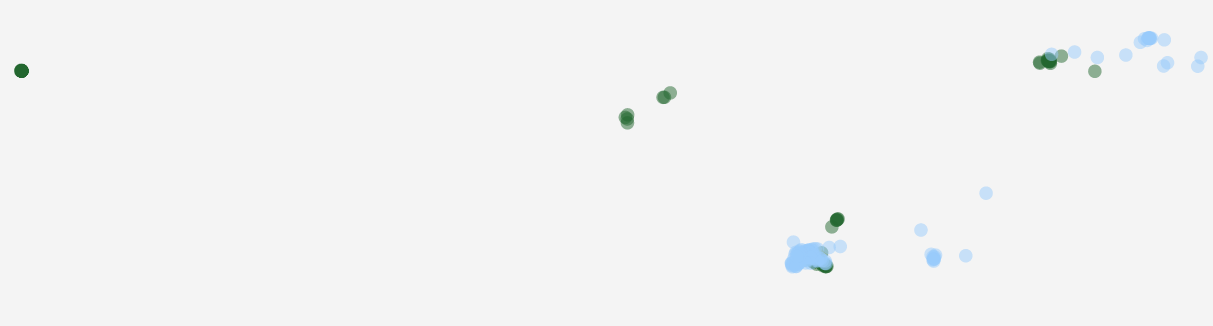}
\caption{Zooming into the two banking77 intents identified in Figure\ref{fig:banking77with2intentscomplexityheatmap}: topup\_transfer\_charge (dark green) and transfer\_fee\_charged (cyan). Each point represents one sentence that is part of the training data. The same issue that our techniques detected on the generated Test data, as exemplified in 
the second row in Table \ref{tab:banking77_sentence_examples} repeats in the training data. There seem to be sub-clusters that include points from both labels. This may be the reason for the high complexity or class confusion.}
\label{fig:banking77with2intentsscatter}
\end{figure}

Tables \ref{tab:banking77_sentence_examples},\ref{tab:citizen_assist_sentence_examples} show example results over the banking77 and CQA datasets, respectively. Our techniques highlight data instances that the classifier failed to classify correctly, yet did so with high confidence about the correctness of the classifications. These instances have high complexity values indicating that one should be more cautious about accepting the classifier's predictions over these instances. It is interesting to notice that high complexity sentences may be short and grammatically correct. Often, they also have no special natural language constructs such as hyphens, contractions and the like. 
For example, the second row in Table \ref{tab:banking77_sentence_examples} shows the sentence `Is there a charge for transfers'. The ground truth label for that sentence is `topup\_transfer\_charge' whereas the predicted label given by the classifier differs and is `transfer\_fee\_charged'. It is unclear what is different between these two intents. We explore these two intents  with our complexity-based visualizations. Notice that the above results are on the augmented Test dataset. We explore these two highlighted intents over the Train dataset. Our results consistently indicate that complexity-related problems in the augmented Test dataset exist also in the original dataset. 
Figure \ref{fig:banking77with2intentscomplexityheatmap} shows a heatmap of the complexity between these intents on the Train dataset. Figure \ref{fig:banking77with2intentsscatter} shows the semantic embedding space of Train dataset examples labeled with these two intents in a reduction from the original high dimensional space into two dimensions. We can pictorially see that these two intents are indeed not well separated. A potential action could be to merge these intents, or to split them differently. 

The examples in Table \ref{tab:banking77_sentence_examples} and in Table \ref{tab:citizen_assist_sentence_examples} demonstrate that it is often difficult to understand why there is a confusion in classification and, thus, also difficult to know a-priori what would be the correct action to solve the issue. Looking, for example, at the first row of Table \ref{tab:banking77_sentence_examples}, it may be the case that this sentence would be a good addition to the training data that would enable the classifier to distinguish between the given and predicted two intents. It could also, however, be the case that these two intents cannot be separated at this stage in the conversation flow. There is a possibility that only after additional information is provided by the user to the chatbot system is it able to handle correctly this query. This requires the chatbot trainer to be aware of the entire system design and its flow. In that case, the handling of the query would be correct even if the classification is incorrect. This is a valid design choice, but it would best be made explicit and added to the Test dataset.

\begin{table*}
  \small
  \centering
    \begin{tabular}{p{7mm}p{1cm}p{2.8cm}p{2.9cm}p{8cm}}
    \toprule
    \textbf{$c$} & \textbf{$h(x,y)$} & \textbf{$y$} & \textbf{$\hat{y}$} & \textbf{Utterance} \\
    \midrule
        0.923 & 2.815 & card\_not\_working & payment\_declined & What's wrong with my card? \newline The clerk at the grocery store said it was declined. \\
        0.780 & 2.707 & topup\_transfer\_charge & transfer\_fee\_charged & Is there a charge for transfers? \\
        0.826 & 2.600 & pending\_top\_up & top\_up\_failed & I tried to top up, but it didn't finish. \\
        0.840 & 0.587 & card\_swallowed & lost\_or\_stolen\_card & The ATM stole my card! \\
        0.865 & 0.818 & wrong\_sum\_cash & cash\_withdrawal\_issue & I got some cash at an ATM earlier, but now app is showing that I withdrew more than I did.  Help! \\
    \bottomrule
    \end{tabular}%
  \captionsetup{width=.95\textwidth}  
  \caption{Utterance examples: Examples of complex utterances from banking77 data-set. Column \textbf{$c$} is the model confidence in the answer, \textbf{$h(x,y)$} is the complexity measure based on the Mahalanobis distance function; \textbf{$y$} and \textbf{$\hat{y}$} are the utterance class and predicted class. Some class names were shortened to fit the limited table space.}
  \label{tab:banking77_sentence_examples}
\end{table*}

\begin{table*}
  \small
  \centering
    \begin{tabular}{p{7mm}p{1cm}p{2.8cm}p{2.9cm}p{8cm}}
    \toprule
    \textbf{$c$} & \textbf{$h(x,y)$} & \textbf{$y$} & \textbf{$\hat{y}$} & \textbf{Utterance} \\
    \midrule
        0.917 & 1.642 & testing\_info & symptoms & I have symptoms. Should I get tested? \\
        0.793 & 2.200 & outside\_activities & avoid\_infection & Is there a charge for transfers? \\
    \bottomrule
    \end{tabular}%
  \captionsetup{width=.95\textwidth}  
  \caption{Utterance examples: Examples of complex utterances from CQA data-set. Column \textbf{$c$} is the model confidence in the answer, \textbf{$h(x,y)$} is the complexity measure based on the Mahalanobis distance function; \textbf{$y$} and \textbf{$\hat{y}$} are the utterance class and predicted class. Some class names were shortened to fit the limited table space.}
  \label{tab:citizen_assist_sentence_examples}%
\end{table*}%










\section{Related work} \label{sec:related}
We provide a methodology and set of technologies for end-to-end testing of chatbot systems.
There exist various commercial and open source technologies for AI testing \cite{seAISurvey,DeepChecksUserGuide,AIEnsured,Infosys,Wipro} and for chatbot testing \cite{Chatbottest,QBox,Coforge,Botium}.
However, all of these do not connect testing goals with business value. Our methodology is business metrics-driven.
Moreover, many of the existing technologies require significant manual effort, e.g., for defining inputs and expected response. Our technologies automate much of the manual effort, such as that required for providing input data, and implement a human-in-the-loop approach that is supported by automatic suggestions.

\section{Conclusions and discussion}
\label{sec:discussion}
We propose a business metrics-driven approach to test and validate the chatbot system.  Our methodology takes a top-down approach to align chatbot testing with business outcomes. Specifically, our approach provide a human-in-the-loop methodology and technologies for maintaining the chatbot intent space and for controlling the behavior of the chatbot following uncontained conversations. Our approach enables quality to be understood and engineered based on the manner in which the test results are analysed and interpreted.  

The work presented in this paper demonstrates the effectiveness and validity of our methodology and technologies in the pre-deployment stage. Further usage by more real world chatbot system testers and designers is needed to further improve the technologies and their integration via the methodology. 

In future work we will address the post-deployment stage, including topic changes and drift detection and remediation. We will build on our large body of work in these areas.  Future work will continue to follow the experiment-first approach and will provide experimental results of putting topic change and drift detection together to show how these may improve the automated detection and suggest actionable insights. We plan to create a framework for allowing the user to maintain the data and explore suggested remediation actions.

In this work we empirically validated the ability to utilize generated realistic data for testing. In future work we will further investigate the properties that make this feasible.

\small {
\bibliography{main} 
}

\end{document}